\documentclass[12pt,preprint]{aastex}

\usepackage{epsfig}
\usepackage{graphicx,color}

\shorttitle{}
\shortauthors{}

\newcommand{\beq}{\begin{equation}}
\newcommand{\eeq}{\end{equation}}

\begin{document}

\title{Magnetic Reconnection with Radiative Cooling. I.~Optically-Thin Regime}


\author{Dmitri A. Uzdensky 
\affil{Center for Integrated Plasma Studies, Physics Department, University of Colorado, Boulder, CO 80309; uzdensky@colorado.edu}
\and Jonathan C.~McKinney
\affil{Department of Physics and Kavli Institute for Particle Astrophysics and Cosmology, Stanford University, Stanford, CA 94305-4060, USA; Chandra Fellow; jmckinne@stanford.edu}}

\begin{abstract}
Magnetic reconnection, a fundamental plasma process associated with a rapid dissipation of magnetic energy, is believed to power  many disruptive phenomena in laboratory plasma devices, the Earth magnetosphere, and the solar corona. Traditional reconnection research, geared towards these rather tenuous environments, has justifiably ignored the effects of radiation on the reconnection process. However, in many reconnecting systems in high-energy astrophysics (e.g., accretion-disk coronae, relativistic jets, magnetar flares) and, potentially, in powerful laser plasma and z-pinch experiments, the energy density is so high that radiation, in particular radiative cooling, may start to play an important role. This observation motivates the development of a theory of high-energy-density radiative magnetic reconnection. As a first step towards this goal, we present in this paper a simple Sweet--Parker-like theory of non-relativistic resistive-MHD reconnection with strong radiative cooling. First, we show how, in the absence of a guide magnetic field, intense cooling leads to a strong compression of the plasma in the reconnection layer, resulting in a higher reconnection rate. The compression ratio and the layer temperature are determined by the balance between ohmic heating and radiative cooling. The lower temperature in the radiatively-cooled layer leads to a higher Spitzer resistivity and hence to an extra enhancement of the reconnection rate. We then apply our general theory to several specific astrophysically important radiative processes (bremsstrahlung, cyclotron, and inverse-Compton) in the optically thin regime, for both the zero-  and strong-guide-field cases. We derive specific expressions for key reconnection parameters, including the reconnection rate. We also discuss the limitations and conditions for applicability of our theory.
\end{abstract}


\keywords{magnetic fields --- magnetic reconnection --- magnetohydrodynamics --- plasmas --- radiation mechanisms: general --- radiation mechanisms: thermal}

\date { \today}

\maketitle


\section{Introduction}

Magnetic reconnection is widely regarded as one of the most important and ubiquitous 
phenomena in plasma physics, with numerous important applications in laboratory plasma physics, 
heliophysics, and astrophysics.
Most of the magnetic reconnection research so far has been driven by our desire to understand magnetic dissipation in various space, solar, and laboratory plasmas, with applications to solar flares, substorms in Earth magnetosphere, and sawtooth disruptions in tokamaks. Importantly, all of these environments are relatively tenuous, low-energy-density, and optically thin, and are adequately described as a collection of non-relativistic charged particles whose numbers are conserved, with no photons.

Magnetic reconnection has also been frequently invoked in a large variety of astrophysical contexts, especially in high-energy astrophysics. Examples include  accretion disks and their coronae and large-scale magnetospheres, jets, gamma-ray bursts (GRBs), pulsar magnetospheres and pulsar winds, flares in soft gamma repeaters (SGRs), etc. Not surprisingly, physical insights obtained from solar-, space- and lab reconnection studies have been often applied to these astrophysical systems. However, it is important to appreciate that the Universe is very diverse and that the range of physical conditions found in various astrophysical environments far exceeds that found within our solar system. In particular, there are some astrophysical phenomena where, on the one hand reconnection has been hypothesized to play an important role and, on the other hand, where the physical parameter regimes are really quite different from those in solar flares, EarthÕs magnetosphere, and laboratory plasmas. Therefore, a straightforward extrapolation of the conventional reconnection scalings to some of these extreme systems is not justified. The necessity to understand how such systems work requires us to develop new theories of magnetic reconnection taking into account several physical processes that are not usually included in conventional reconnection studies.

Some of the most important among these additional physical processes in high-energy astrophysical reconnection are those related to the presence of strong magnetic fields and hence an overall high level of {\it energy density} found in these systems. In particular, at high energy densities various {\it radiative processes} come into play. 
In addition, in some magnetically-dominated environments with low ambient plasma density, e.g., in pulsar magnetospheres and in Active Galactic Nucleus (AGN) jets, one often has to deal with {\it relativistic reconnection}. And finally, at the most extreme end of high energy density astrophysical reconnection --- in environments such as magnetar magnetospheres and central engines and inner jets of Gamma-Ray Bursts --- the energy density is so high, corresponding to radiation temperatures in tens of~keV, or equivalently, magnetic fields of~$10^{12}$~Gauss and more, that, in addition to all the radiative processes mentioned above, {\it pair creation} should become important inside the reconnection layer~\citep{Uzdensky_MacFadyen-2006, Uzdensky-2008}. In the present paper, however, we shall concentrate solely on the role of radiative effects in reconnection and will leave the relativistic effects and pair creation for a future study.

In general, radiation may have several important effects on magnetic reconnection dynamics, 
the most important ones being radiative cooling of the reconnection layer (optically thick or optically thin), 
radiation pressure, and Compton drag (i.e., the radiative resistivity due to collisions between electrons and photons, 
as opposed to electrons and ions as in classical Spitzer resistivity).  
To the best of our knowledge, these aspects of radiative magnetic reconnection have not been adequately explored so far, even though they are critical for several outstanding problems in modern high-energy astrophysics. There is only a handful of references on the subject of reconnection in the presence of radiation \citep{Dorman_Kulsrud-1995, Uzdensky_MacFadyen-2006, Uzdensky-2008, Jaroschek_Hoshino-2009, Nalewajko_etal-2010}, and clearly much more work needs to be done. Thus, there is a clear astrophysical motivation for further effort in this area.

In addition, we believe that this topic should be of strong interest to experimental High-Energy-Density (HED) Physics  --- a new exciting branch of modern physics that emerged in recent years~\citep[e.g.,][]{Drake-2006-book}.  We therefore anticipate a rapid progress in HED reconnection studies facilitated by the advent of new computational and experimental tools and capabilities developed in the HED Physics community, including powerful lasers and Z~pinches. In fact, several HED reconnection experimental studies utilizing laser-produced plasmas with mega-gauss magnetic field have already 
been reported~\citep{Nilson_etal-2006, Li_etal-2007}.


In this paper we are making the first steps towards building up intuition about the role of radiation effects in High-Energy-Density magnetic reconnection. Namely, here we concentrate on the simplest case, corresponding to relatively modest energy densities, where one needs to include only one of the above-mentioned effects --- radiative cooling. In general, prompt radiative cooling in both optically thin and optically thick regimes greatly affects the energy balance and hence the dynamics of the reconnection layer. Different radiation cooling mechanisms may be important in different astrophysical situations, e.g., (1) synchrotron and synchrotron-self-Compton in GRB, AGN, and blazar jets; (2) external inverse-Compton (EIC) cooling of energetic electrons by powerful ambient soft radiation fields in coronae of black holes accreting at a large fraction of the Eddington limit, both in galactic X-ray sources and in AGNs \citep{Goodman_Uzdensky-2008};  (3) radiation diffusion out of an optically-thick pair-dominated reconnection layer in the context of magnetar flares and GRB central engines. In general, we believe that any serious effort in this area will require approaching the reconnection problem as a radiative-transfer problem~\citep{Uzdensky-2008}. For simplicity, however, in this paper we restrict ourselves to the case where the reconnection layer is optically thin. We expect this to be a fair assumption at relatively low energy densities. 
(and also for the case of optically thin neutrino cooling in central engines of long GRBs and core-collapse supernovae, see, e.g., \citealt{Kohri_etal-2005}). The general optically thick case, relevant for higher energy densities, is more complicated because of the above-mentioned need to solve the radiative transfer problem and is therefore left for a future study. We note however, that many of the results derived in this paper, especially in~\S~\ref{sec-general-analysis}, do not depend on the specific properties of the cooling process and therefore are valid for the optically thick case as well.

The outline of our paper is as follows. In~\S~\ref{sec-general-analysis} we lay out the general Sweet--Parker-like theoretical framework for describing compressible antiparallel magnetic reconnection in the presence of strong radiative cooling. In particular, after discussing the main assumptions of our model in~\S~\ref{subsec-assumptions}, we work out an analog of the Sweet--Parker model for the case of strong plasma compression in~\S~\ref{subsec-SP-compressional}, with a specific focus on the case of collisional Spitzer resistivity (\S~\ref{subsec-Spitzer-eta}). Then, in~\S~\ref{subsec-thermodynamics}, we use thermodynamic considerations, namely the heating/cooling balance, to elucidate an important connection between radiative cooling and the plasma compression inside the layer and obtain general expressions for the reconnection layer parameters, including the reconnection rate, in terms of the radiative cooling function. In~\S~\ref{sec-rad-mechanisms}, we apply our general theory to several particularly important radiative mechanisms: the bremsstrahlung (\S~\ref{subsec-bremsstrahlung}) and inverse Compton and cyclotron (\S~\ref{subsec-EIC-Cyclotron}). 
We then devote a substantial part of the paper (\S~\ref{sec-validity}) to discussing various conditions of validity of our assumptions. In particular, in~\S~\ref{subsec-evolution}, we formulate an evolutionary condition for a reconnecting system to be able to reach the strong-compression, strong-cooling regime starting with a very hot initial state and we also investigate thermal stability of the reconnection layer. We demonstrate that bremsstrahlung does not satisfy the evolutionary and stability conditions and thus conclude that, by itself, it cannot lead to the transition to a stable strong-cooling regime, but instead may lead to a cooling catastrophe of the reconnection layer, especially when aided by another process. In the rest of~\S~\ref{sec-validity} we derive the criteria of validity for several of our other assumptions and approximations: collisional resistive magnetohydrodynamics (MHD) (\S~\ref{subsec-collisional}), the optically thin approximation (\S~\ref{subsec-opt-thin}), and the $T_i=T_e$ assumption (\S~\ref{subsec-T_e=T_i}). Finally we consider the strong-guide field case in~\S~\ref{sec-guide}. We summarize our conclusions in~\S~\ref{sec-conclusions}.


\section{General Analysis}
\label{sec-general-analysis}


\subsection{Assumptions}
\label{subsec-assumptions}

In this paper we construct a simple Sweet-Parker-like \citep{Sweet-1958,Parker-1957} model of magnetic reconnection in the presence of relatively strong radiative cooling. Thus, for simplicity, we assume a laminar current sheet configuration involving non-relativistic electron-ion plasma in a steady state. Strictly speaking, we do not expect these assumptions to be satisfied in any real astronomical system with the huge separation of scales (so typical for astrophysics) between the global systems size and the thickness of the reconnection current layer. We are fully aware that, in reality, even in resistive-MHD without externally driven turbulence, very long and thin Sweet--Parker current sheets are likely to be disrupted by the secondary tearing instability~\citep{Loureiro_etal-2007} and to break up into chains of rapidly growing secondary plasmoids~\citep{Samtaney_etal-2009, Bhattacharjee_etal-2009, Cassak_etal-2009}, leading to a very dynamic and non-stationary reconnection process. Correspondingly, we do not expect our laminar model to be straight-forwardly applicable to any concrete astrophysical situation. 
However, since our present work represents a pioneering first effort laying the foundation for a more realistic radiative reconnection theory in the future, we feel that developing a simple Sweet--Parker-like laminar model represents a logical first step. 

Similar to the classical Sweet--Parker analysis, we only consider the two-dimensional (2D) problem with no variation of any quantities in the direction of the current. We shall mostly focus on the case of purely antiparallel reconnection (with no guide field), leaving the more general guide-field case for~\S~\ref{sec-guide}.
In addition, we assume that the plasma is sufficiently dense, and hence collisional enough, that resistive MHD with a (smoothly varying) scalar resistivity is approximately valid. Thus, we ignore the Hall and other extra terms in the generalized Ohm's law, but we do derive the condition for the validity of this assumption. We also ignore the effects of viscosity and thermal conduction.

The key difference between our model and the classical Sweet--Parker model is that we do not assume the plasma to be incompressible and instead of the incompressibility condition we use the balance between ohmic heating and radiative cooling. In this section we develop the general formalism for an arbitrary cooling process and most of the results presented here are independent of its exact nature. We specialize our analysis to the case of optically thin radiative cooling only at the very end of this section. We then consider several specific radiative mechanisms in the next section. 

We would like to remark that a priori (without solving any equations) one should expect a large parameter space for the two main assumptions of our model  --- optically-thin cooling and collisional reconnection  --- to be valid simultaneously. Indeed, it is generally believed that the transition from the collisional to collisionless reconnection takes place when the layer thickness becomes comparable to the ion collisionless skin depth, $d_i = c/\omega_{pi}$. If we are in a situation where we are approaching this transition, the optical depth across the layer is simply $\tau(d_i) = n_e \sigma d_i$, 
where $\sigma$ is the appropriate scattering cross section. Taking for simplicity the Thomson cross-section, 
$\sigma_T=(8\pi/3)\, r_e^2$, where $r_e = e^2/m_ec^2\simeq 2.8\times 10^{-13}\, {\rm cm}$ is the electron classical  radius, we get
$$
\tau(d_i) \sim n_e \sigma_T d_i = {2\over 3}\, \sqrt{m_i\over{m_e}} \, {r_e\over{d_e}} = 
{{4\sqrt{\pi}}\over 3}\,  \sqrt{m_i\over{m_e}} \, (n_e r_e^3)^{1/2} \simeq 100\, (n_e r_e^3)^{1/2} \, ,
$$
where $d_e= c/\omega_{pe}$ is the electron skin-depth. Thus we see that, unless we are dealing with extremely huge densities, approaching~$\sim 10^{-4}\, r_e^{-3}\sim 5\times 10^{33}\, {\rm cm}^{-3}$ (by which time relativistic and/or quantum effects would become important anyway), a reconnection layer on the brink of the transition to collisionless reconnection is unavoidably optically thin to Thomson scattering.

We shall present and analyze the main equations of our model in the next three subsections.
As mentioned above, we basically follow the classical Sweet-Parker analysis but replace the 
incompressibility condition with the full energy equation. In the spirit of the Sweet--Parker model,
we can only do rough estimates, valid only up to factors of order unity. Nevertheless, we shall 
sometimes include such factors in our equations, for future reference.


\subsection{Sweet--Parker Reconnection in the strong-compression limit}
\label{subsec-SP-compressional}

\begin{figure}[t] 
   \centering
   \includegraphics[width=5in]{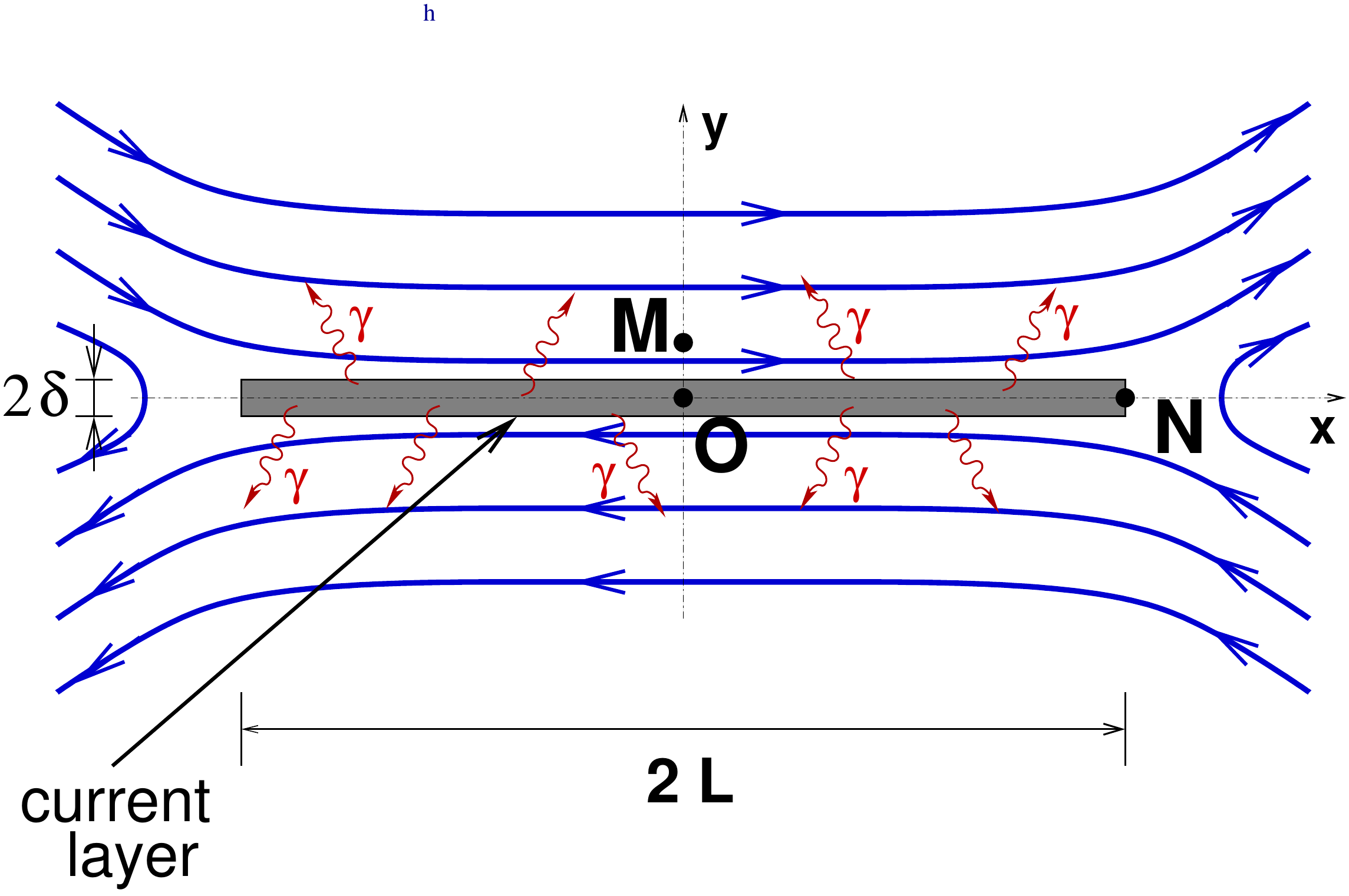} 
   \caption{Reconnection current layer. Magnetic field of strength $B_0$ (shown by blue lines) is reversing across a thin current layer of length~$2L$ and thickness~$2\delta\ll 2 L$. In the presence of radiative cooling, most of the plasma thermal energy (resulting from the ohmic dissipation of the incoming magnetic energy) is carried away across the layer by photons (shown by wiggly red arrows).}
   \label{fig-layer}
\end{figure}

We consider a thin reconnection layer (see Fig.~\ref{fig-layer}) of half-length $L$ and half-thickness~$\delta$.
We denote the direction along the layer as the $x$ direction and the direction across the layer as the $y$~direction.
We denote the inflow velocity into the layer as $v_{\rm rec}$ and the outflow velocity at the end of the layer as~$u$. 
We use the subscript "0" to denote the quantities measured just upstream of the layer, such as the upstream magnetic field~$B_0$, the upstream plasma density~$n_0$, and the upstream temperature~$T_0$. We also introduce the upstream Alfv\'en velocity,
\beq
V_{A0} \equiv {B_0\over{\sqrt{4\pi n_0 m_p}}} \, .
\label{eq-V_A0-def}
\eeq

It will also be convenient to define the global upstream Alfv\'en transit time
\beq
\tau_{A0} \equiv {L\over{V_{A0}}}\, , 
\label{eq-tau_A0-def}
\eeq
and the corresponding Lundquist number
\beq
S_0 \equiv  {{L V_{A0}}\over{\eta}} \gg 1\, ,
\label{eq-S_0-def}
\eeq
where $\eta$ is the magnetic diffusivity.

The quantities at the center of the layer will carry no subscripts. 
We anticipate that strong radiative cooling case considered here is characterized by a strong compression of the plasma inside the layer ($n\gg n_0$), necessary to ensure the pressure balance across it~\citep{Dorman_Kulsrud-1995}. 
To quantify the degree of this compression, we introduce the dimensionless compression ratio 
\beq
A \equiv {n\over{n_0}} \, , 
\label{eq-A-def}
\eeq
which, in the regime of interest (strong cooling), will be expected to be a large number, $A\gg 1$.

The input parameters in our model are the global half-length of the layer, $L$; 
the upstream values of the plasma density ($n_0$), temperature ($T_0$), and
the reconnecting magnetic field ($B_0$); the magnetic diffusivity at the center 
of the layer, $\eta$; and finally the parameters describing the radiative cooling function
(to be discussed below). 

Our primary goal is to compute the 5 main output parameters describing the reconnection layer --- 
such as $n$, $T$, $\delta$, $u$, and $v_{\rm rec}$ (or, equivalently the reconnection electric field 
$E_z= - v_{\rm rec} B_0/c$), in terms of the input parameters. 
We will therefore need 5 algebraic relationships between these parameters that we are going to 
list and discuss in this section. In contrast, in the corresponding  classical incompressible Sweet--Parker analysis, 
one would have only 4 parameters, because the plasma density inside the layer is
fixed to be the same as the upstream density by the incompressibility assumption.

Most, but not all, of these equations will be the same as in the Sweet--Parker model. 
First, we have mass conservation (continuity equation). It is somewhat modified from the traditional Sweet--Parker model to take into account the finite compressibility of the plasma. Assuming that the plasma density everywhere along the midplane of the layer (including at the outflow point~$N$, see Fig.~\ref{fig-layer}) is roughly the same as the density~$n$ at the center of the  layer, we get
\beq
n_0 v_{\rm rec} L  \sim  n u \delta  \quad \Rightarrow \quad  v_{\rm rec} L  \sim  A \, u \delta  \, .
\label{eq-mass-conserv}
\eeq

Next, we consider the set of equations describing the magnetic field: 
Ohm's law, Faraday's law, and Ampere's law. Here, our treatment is quite standard. 
From Faraday's law in a steady state in two dimensions it follows that the out-of-plane
electric field $E_z$ is uniform across the domain; in the ideal-MHD region just outside 
the layer (point M), this field can be written as $E_z = - v_{\rm rec} B_0/c$, whereas at the 
center of the layer point (O), where both ${\bf B}$ and ${\bf v} $ vanish, resistive MHD Ohm's 
law yields $E_z = \eta' j_z$, where $\eta'$ is the plasma resistivity (related to the magnetic diffusivity~$\eta$
via $\eta = \eta' c^2 / 4\pi$). Estimating $j_z$ using Ampere's law as $j_z \simeq - c B_0/4\pi\delta$, 
we get the following important relationship between $v_{\rm rec}$ and~$\delta$: 
\beq
\eta \simeq v_{\rm rec} \delta \, .
\label{eq-Ohm}
\eeq
which is the same as in the Sweet--Parker model.

Furthermore, a quick comparison of the electric field at the inflow point and the ouflow point $N=(x=L, y=0)$, 
allows us to determine the characteristic reconnected magnetic field $B_1\equiv B_y(L,0)$ 
at the end of the reconnection layer: $cE_z = - v_{\rm rec} B_0 = - u B_1$.
(This is basically magnetic flux conservation: the amount of the flux that enters the layer per unit time equals to the amount of flux that leaves the layer; in other words, magnetic flux just reconnects but does not get destroyed). Combining this with the mass conservation equation~(\ref{eq-mass-conserv}), we get an estimate 
\beq
B_1 =  B_y(L,0) = B_0 \, {v_{\rm rec}\over{u}} \sim B_0\, {\delta\over L}\, A \, .
\label{eq-B_1}
\eeq
Note that the resulting expression on the right-hand-side differs by a factor of~$A$
(which we consider to be large) from the corresponding expression in the standard 
incompressible Sweet--Parker theory.
As we shall see shortly, this fact will turn out to be significant when we determine the outflow velocity at the next step.

Now, consider the two components of the equation of motion. 
Since we expect the reconnection inflow velocity into the layer, $v_{\rm rec}$, to be much smaller than the Alfv\'en speed, the $y$ component of the equation of motion simply becomes the pressure balance condition. Neglecting for simplicity the upstream plasma pressure compared with the magnetic pressure (which is  justified by for magnetically-dominant coronal environments), we can then write
\beq
P \equiv P(0,0) = {{B_0^2}\over{8\pi}}   \, .
\label{eq-pressure-balance}
\eeq

Next, we consider the equation of motion along the layer (at the layer's midplane, $y=0$):
\beq
\rho v_x \partial_x v_x = -\, \partial_x P - j_z B_y /c \, .
\label{eq-motion-along}
\eeq
In the original incompressible Sweet--Parker model, the magnetic tension force (the second term on the right-hand side)
turns out to be comparable to the  pressure-gradient force. Then, since one is only interested in getting a rough, order-of-magnitude estimate,  one can drop the magnetic tension term and integrate the resulting equation from $x=0$ to $x=L$, yielding the standard result 
\beq 
{1\over 2} \, \rho u^2  \simeq P(0,0) - P(L,0) \sim P = {{B_0^2}\over{8\pi}} \, .
\label{eq-Bernulli}
\eeq
where we used the result of the pressure balance across the layer in the last step.
This means that the outflow velocity in the Sweet--Parker reconnection is the Alfv\'en velocity computed 
with the upstream reconnecting magnetic field~$B_0$ and the density~$\rho$ (which, we remind the reader, is considered to be uniform in the Sweet--Parker model).  

In our situation, however, this is not the case and the tension force turns out to be much larger than the pressure gradient force. Indeed, using half the value of reconnecting magnetic field~$B_1$ at the exhaust point [see eq.~(\ref{eq-B_1})] to estimate the characteristic reconnected magnetic field inside the layer, we see that the tension force $-j_z B_y/c \sim (B_0/4\pi\delta) (B_0 A \delta/2L) = A\, B_0^2/{8\pi L}$ is by a factor $A\gg 1$ greater than the pressure gradient force $\partial P/\partial x \sim  {B_0^2}/{(8\pi L)}$. 

By equating this tension force with the characteristic magnitude of the inertial term, 
$\rho v_x \partial_x v_x \sim \rho u^2/2L$,  it immediately follows that the outflow velocity at the end of the layer is comparable to the {\it upstream} Alfv\'en velocity (involving the density and the magnetic field just outside of the layer): 
\beq
u \sim {B_0\over{\sqrt{4\pi n_0 m_p}}} =V_{A0} \, .
\label{eq-u}
\eeq

Note that this is different from the usual assumption (which, we claim, is not justified) that 
the outflow velocity in the compressible case is that corresponding to the {\it composite} Alfv\'en velocity
defined with the compressed density inside the layer and the magnetic field outside the layer 
\citep{Parker-1963} --- a result that one would have obtained by taking into account the pressure gradient force only.

Once the compression ratio~$A$ is found, solving the rest of the problem (i.e., finding the reconnection rate) 
is rather straight-forward, especially since the remaining equations that we need to complete this task [namely, eqs.~(\ref{eq-mass-conserv})-(\ref{eq-Ohm}) and~(\ref{eq-u})] are the same as in the classical Sweet-Parker analysis.
First, combining Ohm's law~(\ref{eq-Ohm}) with the continuity equation~(\ref{eq-mass-conserv}), we get 
$ v_{\rm rec} L \sim A u \delta \simeq u\, (\eta/v_{\rm rec})\, A $. Then, substituting $u\sim V_{A0}$ from equation~(\ref{eq-u}), we immediately arrive at
\beq
v_{\rm rec}^2  \sim {A\eta\over{L}}\, u \sim {{\eta V_{A0}}\over{L}}\, A \, ,
\label{eq-v_rec-A}
\eeq
which can be rewritten in terms of the Lundquist number~(\ref{eq-S_0-def}) as
\beq
v_{\rm rec} \sim V_{A0}\, S_0^{-1/2}\, A^{1/2} \, .
\label{eq-v_rec-A2}
\eeq
This means that compression makes reconnection faster by a factor~$A^{1/2}$, 
compared with the incompressible Sweet--Parker result.  Finally, the thickness of the layer is given by
\beq
\delta \sim {\eta\over{v_{\rm rec}}} \sim L \,  S_0^{-1/2}\, A^{-1/2} = \delta_{\rm SP}\, A^{-1/2} \, .
\label{eq-delta}
\eeq
That is, the compressed layer is thinner than the corresponding incompressible Sweet--Parker layer by the same factor of~$A^{1/2}$.


\subsection{Strong-Compression Reconnection Regime with Spitzer Resistivity}
\label{subsec-Spitzer-eta}

Next, it is interesting to express the above reconnection layer parameters using the actual expression for the resistivity. Assuming that the dominant source of resistivity is the perpendicular Spitzer resistivity due to electron-ion collisions (we shall leave the case of Compton-drag resistivity, important in many astrophysical situations, for a future study), we can write the corresponding magnetic diffusivity as~\citep{Spitzer-1962, Braginskii-1965}
\beq
\eta_{\perp} = C_\eta \, c r_e\,  \ln\Lambda \, \theta_e^{-3/2} \, , 
\label{eq-eta_Spitzer}
\eeq
where $C_\eta = \sqrt{2}/3\sqrt{\pi} \simeq 0.27$ and $\ln\Lambda$ is the Coulomb logarithm. Then, it is straight-forward to obtain the following expression for the Lundquist number as a function of density~$n_e$ and temperature~$\theta_e$:
\beq
S(n,\theta_e) = 
{L V_{A}\over{\eta_\perp}} =  C_\eta^{-1}\, {L\over{r_e \ln\Lambda}}\, {V_{A}\over c}\, \theta_e^{3/2} \, .
\label{eq-Spitzer-S}
\eeq
What we need in our theory [see eqs.~(\ref{eq-v_rec-A2})-(\ref{eq-delta})], is the Lundquist number defined with the actual central electron temperature~$\theta_e$ (which comes in via the resistivity) but with the upstream density~$n_0$. Furthermore, we would like to express this Lundquist number as a function of the compression ratio~$A$. 
To do this, we make use of the pressure balance equation~(\ref{eq-pressure-balance}) written as
\beq
{{B_0^2}\over{8\pi}}  =  P = 2 n k_B T \, ,
\label{eq-pressure-balance-T}
\eeq
where the factor of 2 reflects the two equal contributions from electrons and ions (whose temperatures are assumed here to be the same, see~\S~\ref{subsec-T_e=T_i}). From this, we can express the central layer temperature in terms of the compression ratio~$A$:
\beq
k_B T(A) = {{B_0^2}\over{16\pi n}}  = A^{-1} \, k_B T_{\rm eq} \, ,
\label{eq-T_of_A}
\eeq
where $k_B T_{\rm eq}\equiv B_0^2/16\pi n_0$ corresponds to the central temperature that the layer would need to have without compression (i.e., for $A=1$). This temperature can also be written in a convenient dimensionless form as
\beq
\theta_e \equiv {{k_B T}\over{m_e c^2}} = A^{-1}\, {{B_0^2}\over{16\pi n_0 m_e c^2}} = 
{1\over{4A}}\, {m_p\over{m_e}}\, {{V_{A0}^2}\over{c^2}} \, .  
\label{eq-theta_e}
\eeq

Substituting this expression for the temperature into equation~(\ref{eq-Spitzer-S}), we get
\beq
S_0 = S(n_0,\theta_e) = {{L V_{A0}}\over{\eta_{\rm Sp}}} =
C_\eta^{-1}\, {L\over{r_e \ln\Lambda}}\, {V_{A0}\over c}\, \theta_e^{3/2} =  
{1\over{8\,C_\eta}}\,  {L\over{r_e \ln\Lambda}}\, \biggl({V_{A0}\over c}\biggr)^4 \, \biggl({{m_p}\over{m_e}}\biggr)^{3/2}\, A^{-3/2} \, .
\eeq

Correspondingly, the reconnection velocity~(\ref{eq-v_rec-A2}) is given by
\beq
{{v_{\rm rec}}\over c} \sim  \sqrt{{8 C_\eta\,  r_e \ln\Lambda}\over{L}} \, 
\biggl({V_{A0}\over{c}} \biggl)^{-1}\, \biggl({{m_e}\over{m_p}}\biggr)^{3/4}\, A^{5/4} \, ,
 \label{eq-v_rec_of_A}
\eeq
and the the thickness of the layer~(\ref{eq-delta}) becomes
\beq
\delta^2 \sim {{L^2}\over{S_0 A}} \sim C_\eta\,(L r_e) \, \ln\Lambda \, {c\over{V_{A0}}}\, A^{-1}\, \theta_e^{-3/2} = 8 C_\eta\, (L r_e)\, \ln\Lambda\, \biggl({V_{A0}\over{c}} \biggl)^{-4}\, \biggl({{m_e}\over{m_p}}\biggr)^{3/2}\, A^{1/2}\, .
 \label{eq-delta_of_A}
\eeq


\subsection{Thermodynamics of the Radiativelly-Cooled Reconnection Layer and the Compression Ratio}
\label{subsec-thermodynamics}

Finally, we need to consider one more equation that would enable us to determine the compression ratio~$A$. 
To do this, we will need to invoke thermodynamic considerations. As we shall see, this part of the problem becomes one-dimensional (1D) in the strong cooling limit and effectively decouples from the rest of the analysis.
Naturally, one wants to look at the energy conservation equation. This equation is usually not needed
in the classical (incompressible) Sweet--Parker analysis, because there it is replaced by the incompressibility 
condition (although the effects of plasma compressibility have been discussed in the literature, 
see, e.g., \citealt{Parker-1957, Parker-1963}). But since here we are interested in a plasma that is strongly compressible, 
we have to include this equation in our analysis. We note that, even though here we are mostly interested in the optically thin case, most of our results in this subsection apply, or can be easily extended, to the general optically thick case where one has to solve a 1D radiative transfer problem.

In general, energy conservation in a steady state can be viewed as a balance between 
the flux of magnetic enthalpy flowing onto the layer (for simplicity, we here neglect the upstream 
thermal enthalpy compared with the magnetic one; that is, in this paper we restrict ourselves to 
the case of low $\beta_{\rm upstream}$ plasma; however, our analysis can be straight-forwardly 
generalized to the finite-$\beta_{\rm upstream}$ case) on the one hand and, on the other, 
the sum of the advection of plasma thermal enthalpy and kinetic energy  out of the layer, 
plus (a new element!) radiative cooling across the layer. The magnetic enthalpy flux per 
unit surface area is the sum of the inflow of magnetic energy, $v_{\rm rec} B_0^2/8\pi$, 
and the work done by the magnetic pressure, which is also~$v_{\rm rec} B_0^2/8\pi$;
thus, the total magnetic enthalpy flux is $v_{\rm rec} B_0^2/4\pi$, which of course is the same 
as the Poynting flux, $S_{\rm Poynt} = cE B_0/4\pi = v_{\rm rec} B_0^2/4\pi$ (we use the subscript
``Poynt" to avoid confusion between the Poynting flux and the Lundquist number~$S$).

Denoting the advective and radiative energy fluxes by $F_{\rm adv}$, and $F_{\rm rad}$, respectively, we can write
the energy conservation condition (from the Eulerian viewpoint) as
\beq
S_{\rm Poynt} \, L \sim  {c\over{4\pi}}\, E B_0 \, L \sim  F_{\rm adv} \delta + F_{\rm rad} L \, .
\label{eq-energy-conserv-general}
\eeq

The total advective energy flux out of the layer can be estimated as
\beq
F_{\rm adv} \delta \sim u\, (\rho u^2/2 + 5 nk_BT)\, \delta \, , 
\label{eq-F_adv-1}
\eeq
where the first term is the flux of kinetic energy and the second is the flux of thermal enthalpy of a two-species gas.
As we showed above, the condition of pressure balance across the layer dictates that $2nk_BT$ be equal, or at least comparable, 
to~$B_0^2/8\pi$.
At the same time, the equation of motion along the layer dictates that the plasma kinetic energy is $\rho u^2/2 \sim
A B_0^2/8\pi$. Thus, although the two energy fluxes are comparable in the incompressible Sweet--Parker case, in the strong-compression $A\gg 1$ case, the kinetic energy flux always dominates over thermal enthalpy flux . It is then easy to see that in this case
the ratio of the total advective energy flux out of the layer to the total Poynting flux into the layer can be estimated as
\beq
{{F_{\rm adv} \delta}\over{S_{\rm Poynt}\,L}} \sim \biggl({{uA \delta}\over{v_{\rm rec} L}}\biggr) \, .
\label{eq-F_adv-2}
\eeq

Using the continuity equation~(\ref{eq-mass-conserv}), we thus  immediately see that  
\beq
{{F_{\rm adv} \delta}\over{S_{\rm Poynt}\,L}} = O(1) \, . 
\label{eq-F_adv-3}
\eeq
This is a noteworthy result. It means that, independent of the strength of radiative cooling and hence of plasma compression,
a finite fraction of the incoming magnetic energy is not dissipated into heat (which is then promptly radiated away), but instead is converted directly into the mechanical kinetic energy of the outflow.
This implies that, strictly speaking, one cannot use the total energy conservation to estimate the compression ratio~$A$ by simply equating the Poynting flux with the radiative losses!
What one should do instead, is to look at the fate of plasma {\it entropy}, i.e., at the balance between the plasma heating due to the dissipation of magnetic energy (the ohmic heating) and the plasma cooling both via the advection of heat out of the layer and via the radiation losses.
In other words, one needs to focus on the thermal content of the plasma, as opposed to its total energy.
Then one can say that the system is in the strong radiative cooling regime if the ohmic heating rate is primarily 
balanced by radiative losses, whereas the heat losses by advection are small. 
This is in contrast with the classical (non-radiative) Sweet--Parker model, where it is straight-forward to show that
ohmic heating is balanced by the advection of thermal energy out of the layer. 

The ohmic heating rate in our model can be estimated as
\beq
Q_{\rm ohm} = \eta' j^2 \sim \eta' \biggl({c B_0\over{4\pi\delta}}\biggr)^2 = 
{\eta\over{\delta^2}} \, {{B_0^2}\over{4\pi }}  \sim {{v_{\rm rec}}\over\delta}\, {{B_0^2}\over{4\pi}} \, ,
\label{eq-Q_ohm-1}
\eeq
where $\eta' = 4\pi\,\eta/c^2$ is the resistivity, and where we made use of equation~(\ref{eq-Ohm}).
Combining this with the mass conservation condition~(\ref{eq-mass-conserv}), $v_{\rm rec} L \sim u\delta A$, 
and with~(\ref{eq-u}), we find
\beq
Q_{\rm ohm}  \sim A\, {{B_0^2}\over{4\pi \tau_{A0}}} \equiv A\, Q_0 \, ,
\label{eq-Q_ohm-2}
\eeq
where we defined a characteristic power per unit volume $Q_0\equiv B_0^2/(4\pi\tau_{A0})$. 

On the other hand, the heat loss rate due to the advection out of the layer is, roughly, just the thermal energy density 
divided by the advection time along the layer, $L/u \sim L/V_{A0} = \tau_{A0}$:
\beq
Q_{\rm adv} \sim { 3\, n k_B T\over{L/u}} \sim  {{B_0^2}\over{4\pi \tau_{A0}}} = Q_0 \sim Q_{\rm ohm}/A \, ,
\label{eq-Q_adv}
\eeq
where we ignored factors of order unity.

Thus, not surprisingly, the condition that most of the heat resulting from magnetic energy dissipation is promptly radiated away (instead of being advected with the plasma outflow) is equivalent to the condition of strong compression ($A\gg 1$, $n\gg n_0$), which is what we generally assume in this paper. The exact criterion of when this is the case depends on the specific model for radiative cooling; we will consider this in more detail in~\S~\ref{sec-rad-mechanisms}.

In general, one can say that the reconnection process partly converts magnetic energy into the bulk kinetic energy of the outflow (by performing mechanical work) and partly dissipates it into heat (by ohmic heating). In both the Sweet--Parker regime and the radiative regime considered here the two parts are roughly comparable. However, whereas in the Sweet--Parker case both of these forms of plasma energy are taken out along the layer by the plasma outflow, in the strong radiative cooling case the kinetic part of the energy is still advected out along the layer but most of the thermal energy is promptly radiated away.

Provided that we can neglect the advective heat loss compared with the radiative loss, 
we can now write the heating-cooling balance in the optically-thin regime as 
\beq
Q_{\rm rad}(n,T)  = Q_{\rm ohm} \sim A\, Q_0    \, , 
\label{eq-heat-cool-1}
\eeq
where the functional form of the volumetric radiative cooling rate $Q_{\rm rad}$ on $n$, and $T$,
(in general it may also depend on~$B$, e.g., in the case of cyclotron cooling) depends on the specific radiative mechanism. Assuming that this functional dependence is known, this equation yields the desired algebraic 
equation for~$A$:
\beq
Q_{\rm rad} [n(A),T(A)]  \simeq A \, Q_0 = A\, {{B_0^2}\over{4\pi \tau_{A0}}}  \, .
\label{eq-entropy-A}
\eeq
Here, the arguments $n$, $T$ of the radiative cooling function on the left-hand-side are viewed as
functions of~$A$ and of the input plasma parameters~$n_0$, and~$B_0$, given by $n(A) = n_0 A$ 
and equation~(\ref{eq-T_of_A}), respectively.

It is very important to note that the above equation~(\ref{eq-entropy-A}) that determines~$A$ does not explicitly 
contain the resistivity~$\eta$; it only involves the input plasma parameters $n_0$ and~$B_0$ and the length~$L$ 
that are considered to be given. This means that the one-dimensional (1D) problem of determining the thermal structure of the layer (i.e., determining the density~$n$ and temperature~$T$ inside the layer) essentially decouples from the 2D reconnection problem itself. This enables us to find the compression ratio~$A$ and hence~$n$ and~$T$ independent of the resistivity and the reconnection rate. It is interesting to note that, in general, the resulting layer temperature may turn out to be higher or lower than the ambient (upstream) plasma temperature.


\section{Specific Radiative Mechanisms}
\label{sec-rad-mechanisms}


To make further progress, we need to make a specific choice of the dominant radiative cooling mechanism. 
We shall consider several plausible processes in the following subsections.


\subsection{Bremsstrahlung and Atomic-line Cooling}
\label{subsec-bremsstrahlung} 
 
In many astrophysical environments, e.g., from the solar corona to the hot gas in galaxy clusters,
the predominant cooling process is free-free cooling due to binary collisions between electrons and ions. 
Because it is collisional in nature, the cooling rate is proportional to the square of the plasma density, i.e., 
\beq
Q_{\rm rad}(n,T) = n^2 \, \Lambda(T) \, .
\label{eq-Q_ff}
\eeq
where $\Lambda(T)$ is the so-called cooling function. 
Correspondingly, equation (\ref{eq-entropy-A}) can be written as
\beq
 A\, \Lambda(T) \simeq  Q_0\, n_0^{-2} = {{B_0^2}\over{4\pi\tau_{A0}}} \, n_0^{-2} \, . 
 \label{eq-energy-ff}
\eeq

In general, for a plasma with a non-trivial (e.g., solar) chemical composition, the functional form 
of $\Lambda(T)$ is quite complicated. In particular, at temperatures between $10^5$~K and $10^7$~K 
cooling is dominated by various strong atomic lines (mostly those of carbon, oxygen, and iron) and looks 
like a mountainous landscape; in solar physics it is usually represented by a broken power-law~\citep[e.g.,][]{Raymond_Smith-1977}.
However, above about $10^7$~K atomic lines mostly disappear and the cooling function 
is dominated by a simple bremsstrahlung cooling~\citep{RL-1979}:
\beq
\Lambda(T) = \kappa\, T^{1/2} = \kappa\, {{B_0}\over{\sqrt{16\pi n_0 k_B}}} \,  A^{-1/2} =
\kappa \, V_{A0}\, (m_p/4k_B)^{1/2}\, A^{-1/2} \, .
\label{eq-Lambda-brems}
\eeq
where for a nonrelativistic Maxwellian hydrogen plasma, 
\beq
\kappa = {{16\sqrt{2\pi}}\over{3\sqrt{3}}} \, c^2 r_e^2\, \sqrt{m_e k_B}\, \alpha_{\rm fs}\, \bar{g}_{\rm ff} 
\simeq 1.4 \times 10^{-27} \, \bar{g}_{\rm ff} \, {\rm erg}\, {\rm cm}^{3}\, {\rm s}^{-1} \, {\rm K}^{-1/2}\, .
\label{eq-kappa-brems}
\eeq
where $\alpha_{\rm fs} = e^2/\hbar c\simeq 1/137$ is the fine structure constant and $\bar{g}_{\rm ff}$ is the frequency averaged free-free Gaunt factor~\citep{RL-1979}. Since here we are interested in the regime where bremsstrahlung dominates over atomic line emission and where, at the same time, the electrons are nonrelativistic, we have to restrict our consideration to the temperature range of $10^7\ {\rm K} \lesssim T \ll  5\times 10^9\ {\rm K}$; in this range the averaged free-free Gaunt factor varies approximately between~1.1 and~1.2 \citep{Karzas_Latter-1961}. 

Thus, the bremsstrahlung cooling rate can be expressed as a function of~$A$ as
\beq
Q_{\rm rad}^{brems}(A) = 
\sqrt{2\over{3\pi}}\, \alpha_{\rm fs}\, \tau_T\,\bar{g}_{\rm ff}\, {{n_0 c^2\sqrt{m_e m_p}}\over{\tau_{A0}}} \, A^{3/2} = 
\sqrt{2\over{3\pi}}\, \alpha_{\rm fs}\,\tau_T\, \bar{g}_{\rm ff}\, \sqrt{m_e\over{m_p}}\, {c^2\over{V_{A0}^2}}\,Q_0\,A^{3/2}\,, 
\label{eq-Q_ff-A}
\eeq
where 
\beq 
\tau_T = \sigma_T\, n_0 \, L = {8\pi\over 3}\, r_e^2 \, n_0 \, L 
\label{eq-def-tau_T}
\eeq
is the upstream-density Thomson optical depth {\it along} the layer.

Because the above functional form of the cooling function for pure Bremsstrahlung is particularly simple, 
we can solve equation~(\ref{eq-energy-ff}) corresponding to this case explicitly. We have
\beq
A^{1/2}  \simeq  2\, {{V_{A0}^2}\over{n_0 L}}\,  \kappa^{-1}\, (k_B /m_p)^{1/2}   
 \simeq  \sqrt{{3\pi}\over{2}}\,\bar{g}_{\rm ff}^{-1}\, {V_{A0}^2\over{c^2}}\, \sqrt{m_p\over{m_e}}  \, (\alpha_{\rm fs} \tau_T)^{-1} \simeq 10^{30} \,  {{B_0^2}\over{n_0^2 \, L}}\  {\rm in\ cgs\ units} \, .
\label{eq-brems-A}
\eeq
 
Then, the condition for the existence of an $A\gg 1$ solution, corresponding to the strong-cooling, strong-compression regime, can be cast as 
\beq
 {V_{A0}^2\over{c^2}} \gg  
0.5\, \alpha_{\rm fs} \,  \tau_T \, \sqrt{m_e\over{m_p}} \simeq  10^{-4}\, \tau_T \, , 
\label{eq-brems-condition}
 \eeq 
or, $B_0^2/n_0^2\, L  \gg 10^{-30}$ in cgs units.

If this condition is satisfied, we can use equations~(\ref{eq-v_rec_of_A}) and~(\ref{eq-delta_of_A}) to determine the other key parameters of the reconnection layer 
\begin{eqnarray}
{v_{\rm rec}\over{c}} &\sim &  
(3\pi)^{5/4}\, (\sqrt{2}\,C_\eta)^{1/2} \, \bar{g}_{\rm ff}^{-5/2}\, \sqrt{{r_e\ln\Lambda}\over{L}} \, 
\biggl({{V_{A0}}\over c}\biggr)^4\, \biggl({m_p\over{m_e}}\biggr)^{1/2} \, (\alpha_{\rm fs} \tau_T)^{-5/2}  \nonumber \\
&\simeq &  10\, \bar{g}_{\rm ff}^{-5/2}\, \sqrt{{r_e\ln\Lambda}\over{L}} \, 
\biggl({{V_{A0}}\over c}\biggr)^4\, \biggl({m_p\over{m_e}}\biggr)^{1/2} \, (\alpha_{\rm fs} \tau_T)^{-5/2} \, , \\
\delta^2 &\sim & 3\,\sqrt{3\over{2\pi}}\, C_\eta\, \bar{g}_{\rm ff}^{-1}\, \alpha_{\rm fs}^{-1}\, {{\ln\Lambda}\over{n_0\, r_e}} \, {m_e\over{m_p}}\, \biggl({{V_{A0}}\over c}\biggr)^{-2} \nonumber \\
&\simeq &  {\sqrt{3}\over{\pi}}\, \bar{g}_{\rm ff}^{-1}\, \alpha_{\rm fs}^{-1}\, {{\ln\Lambda}\over{n_0\, r_e}} \, {m_e\over{m_p}}\, \biggl({{V_{A0}}\over c}\biggr)^{-2}  \, .
\end{eqnarray}

Interestingly, we see that the thickness of the layer becomes independent of the system size~$L$ in this case; it can be conveniently expressed in terms of the upstream electron collisionless skin depth~$d_{e0}^2 = m_e c^2/4\pi n_0 e^2=  (4\pi n_0 r_e)^{-1}$ as
\beq
{{\delta}\over{d_{e0}}} \simeq 
2\,\cdot 3^{1/4}\, \sqrt{{\ln\Lambda}\over{\bar{g}_{\rm ff}\, \alpha_{\rm fs}}}\,  {c\over{V_{Ae}}} \, , 
\label{eq-delta-d_e-bremsstrahlung}
\eeq
where $V_{Ae} = B_0/(4\pi n_0 m_e)^{1/2}$ is the electron Alfv\'en velocity.  


\subsection{External Inverse Compton and Cyclotron Cooling} 
\label{subsec-EIC-Cyclotron}

External Inverse Compton (EIC) and cyclo/synchrotron cooling are two of the most important radiative cooling mechanisms in High-Energy Astrophysics, playing a dominant role in regulating plasma temperatures in such environments as coronae of  accreting black-holes and in astrophysical jets, etc. 
For example, if  the plasma is immersed in an isotropic external soft radiation field with energy density~$U_{\rm rad}$, then inverse Compton scattering can an important cooling mechanism, provided that the characteristic energy of external photons is  less than the temperature of the layer. This is in fact the case for accretion disk coronae of powerful black holes accreting at a sizable fraction of the Eddington limit, e.g., in quasars. In these systems, the relatively cool accretion disk provides a powerful source of soft photons that very effectively cool the electrons accelerated in coronal reconnection events. Likewise, when the plasma has a significant relativistic electron or electron-positron component, as is believed to be the case in, e.g.,  pulsar wind nebulae and in AGN and GRB jets, and is immersed in a strong magnetic field,  then synchrotron emission may provide the dominant cooling mechanism. However, since in the present paper we, for simplicity, focus on non-relativistic plasmas where synchrotron emission is absent, we cannot consider this mechanism here and we shall therefore leave it to a future study. As for its non-relativistic analog, the cyclotron cooling, it turns out that for the non-relativistic zero-guide-field case considered in the main part of this paper, it is not effective. The reason for this is the following. By virtue of the pressure balance condition~(\ref{eq-pressure-balance-T}), the fundamental electron cyclotron frequency, $\Omega_{ce} = eB/m_e c$ , can be expressed in terms of the layer's plasma frequency and the normalized electron temperature as $\Omega_{ce} = 2\omega_{pe}\, \theta_e^{1/2}$. Thus, since we assume $\theta_e \ll 1$, we see that $\Omega_{ce} < \omega_{pe}$ inside the layer; therefore, waves at the electron cyclotron frequency cannot propagate across the magnetic field and hence across the layer. There may still be some cooling via higher electron cyclotron harmonics, but we neglect it here. 
We note, however, that in the presence of a strong guide field the above pressure balance considerations do not apply and one may have a situation where $\Omega_{ce} > \omega_{pe}$ and hence where cyclotron cooling can be effective. 
We shall consider this situation in~\S~\ref{sec-guide} and devote the rest of this section to inverse-Compton cooling.
At the same time, we still would like to remark that, even though they are beyond the scope of the present paper, the effects of cyclo/syncrotron cooling that one would have in the case of a relativistically hot plasma can be treated in a very similar manner.

Let us investigate the effect of EIC cooling on magnetic reconnection in more detail. The optically-thin radiative cooling rate of a single electron can be written as~\citep{RL-1979}
\beq
{{Q_{\rm rad}}\over{n}} = {4\over 3}\, \sigma_T\, c \, U \, \beta^2 \gamma^2 \, , 
\eeq
where $U$ is the soft radiation energy density, $U=U_{\rm rad}$. 
(In the case of cyclotron/synchrotron cooling, a similar expression would apply with $U$ being the magnetic energy density, $U_{\rm mag}$, and in the general case when both of these processes are active, $U$ should be the sum 
of the two energy densities, $U=U_{\rm rad}+U_{\rm mag}$.) 

Next, since in this paper we are focusing on a non-relativistic ($\beta\ll 1$) Maxwellian plasma with the average electron kinetic energy of $<m_e c^2 \beta^2> = 3/2 \, k_B T$, we can write the radiative cooling rate per unit volume as
\beq
Q_{\rm rad} = \sigma_T\, c \, U \, {2nk_B T\over{m_e c^2}} = \sigma_T\, c \, U \, {P\over{m_e c^2}} \, , 
\label{eq-Q_rad-EIC-cyclo}
\eeq
where we assumed $T_i=T_e=T$ to make the last step. Then, using the pressure balance condition $P =  B_0^2/8\pi$, we get 
\beq
Q_{\rm rad} =  \sigma_T\, c \, U \, {{B_0^2}\over{8\pi\, m_e c^2}} = 
{\tau_T\over 2}\, {c\over{V_{A0}}}\, {U\over{n_0 m_e c^2}} \, Q_0\,  .
\label{eq-Q_rad-EIC-cyclo-2}
\eeq
Interestingly, we see that the volumetric cooling rate is independent of the compression ratio~$A$. 
Then, substituting this expression into equation~(\ref{eq-entropy-A}), we can immediately read off the compression ratio:
\beq
A \sim {{Q_{\rm rad}}\over{Q_0}} = \sigma_T\, c \, {{U \tau_{A0}}\over{2\, m_e c^2}}  = 
{\tau_T\over 2}\, {c\over{V_{A0}}}\, {U\over{n_0 m_e c^2}} =
{\tau_T\over 4}\, {{V_{A0}}\over c}\, {m_p\over{m_e}}\, {{8\pi\, U}\over{B_0^2}} \, .
\label{eq-EIC-A}
\eeq

Correspondingly, the normalized layer temperature~(\ref{eq-theta_e}) is 
\beq
\theta_e = A^{-1}\, {B_0^2\over{16\pi \, n_0 m_e c^2}} \sim  
{{B_0^2}\over{8\pi\, U}} \, \tau_T^{-1}\, {{V_{A0}}\over c} \, .
\label{eq-EIC-theta}
\eeq

Then, the condition that $A\gg 1$ is equivalent to the condition that the layer is sufficiently long:
\beq
\tau_T \gg  2\, {V_{A0}\over c}\, {{n_0 m_e c^2}\over{U}} =
4\, {m_e\over{m_p}} \, {c\over{V_{A0}}}\, {{B_0^2}\over{8\pi\, U}}\, , 
\eeq
and the condition that we are dealing with a non-relativistic plasma, $\theta_e \ll 1$, as we have assumed, 
becomes 
\beq
{{8\pi\, U}\over{B_0^2}}\, \tau_T \gg  {V_{A0}\over{c}} \, .
\eeq
By combining these two inequalities, we get the following necessary condition:
\beq
{{8\pi\, U}\over{B_0^2}}\, \tau_T > 2\, \sqrt{m_e\over{m_p}} \simeq 0.05 \, .
\eeq

Finally, the reconnection layer parameters are:
\begin{eqnarray}
{{v_{\rm rec}}\over c} &\sim & {\sqrt{C_\eta}\over{2}}\, \tau_T^{5/4}\, \sqrt{{r_e\over L}\, \ln\Lambda}\, 
\biggl({{V_{A0}}\over c}\biggr)^{1/4}\, \sqrt{m_p\over{m_e}}\, \biggl({{8\pi\, U}\over{B_0^2}}\biggr)^{5/4}\, , \\
\delta^2 &\sim & 4 C_\eta\, L r_e \ln\Lambda\, \tau_T^{1/2}\, \biggl({{V_{A0}}\over{c}}\biggr)^{-7/2}\, 
{m_e\over{m_p}}\, \biggl({{8\pi\, U}\over{B_0^2}}\biggr)^{1/2} \, .
\end{eqnarray}

%


\section{Additional Validity Conditions}
\label{sec-validity}

Let us now discuss the conditions required for the strong-cooling model developed in the preceding sections  to be valid. 


\subsection{The Evolutionary Condition and Thermal Stability of the Reconnection Layer}
\label{subsec-evolution}

First, an obvious necessary condition for the strong-cooling, strong-compression regime is that equation~(\ref{eq-entropy-A}) has a large-$A$ solution, $A\gg 1$. However, the actual situation is somewhat more subtle. The condition $A\gg 1$ 
is just the condition for the {\it existence} of a stationary strongly-cooled state of the reconnection layer. In addition, however, we must impose an extra evolutionary condition for the system to be able to reach this state. As we shall see below, this will result in a certain requirement for the radiative cooling function. The picture that we have in mind here is the following. The ambient plasma upstream of the reconnection layer is rather tenuous; when it just enters the layer, it becomes subject to ohmic heating and its temperature rises, whereas its density does not change appreciably at first. If radiative cooling can be neglected, one always gets the classical Sweet--Parker layer solution, with relatively low density $n\simeq n_0$ and relatively high temperature~$T\simeq T_{\rm eq}$ (corresponding to $A\simeq 1$). The transition to the strong-cooling, strong-compression $A\gg 1$ regime described in the previous sections happens only if that $A\simeq 1$ Sweet--Parker layer becomes unsustainable in the presence of radiative cooling, i.e., if it is able to cool and collapse towards the $A\gg 1$ solution sufficiently rapidly. For this to happen, we must require that the radiative cooling of the corresponding $A\simeq 1$ Sweet--Parker solution be stronger that the corresponding Ohmic heating, i.e.:
\beq
Q_{\rm rad}[n_0, T_{\rm eq}(n_0)] > Q_{\rm ohm}[n_0, T_{\rm eq}(n_0)]  \simeq Q_0\, ,
\label{eq-SP-cooling-condition}
\eeq
where we used equation~(\ref{eq-Q_ohm-2}) in the last step.

Now, assuming that a stationary strong-cooling solution with $A\gg 1$ does exist, it is convenient to make use of the corresponding heating-cooling balance equation $Q_{\rm rad}(A) = Q_{\rm ohm}(A) \simeq A Q_0$ [see eq.~(\ref{eq-heat-cool-1})] and recast the above condition in a form that involves only the cooling rate function and the value of~$A$ corresponding to the strong-cooling solution:
\beq
Q_{\rm rad}[n_0, T_{\rm eq}(n_0)] > Q_{\rm ohm}[n_0, T_{\rm eq}(n_0)]  \simeq Q_0 \simeq 
Q_{\rm ohm}(A)/A = Q_{\rm rad}(A)/A\, , 
\eeq
i.e., 
\beq
Q_{\rm rad} [n(A),T(A)]  < A\, Q_{\rm rad}[n_0, T_{\rm eq}(n_0)] \, , \qquad A\gg 1 \, .
\label{eq-evolutionary-condition}
\eeq

We thus see that the necessary requirement that a reconnecting layer is able to evolve to the strong-compression, strong-cooling regime, imposes a certain nontrivial constraint on the {\it functional form} of the volumetric radiative cooling rate~$Q(n,T)$. This means that not every radiative process is actually able to cause the system to transition to the strong-cooling regime, large-$A$ solution of equation~(\ref{eq-entropy-A}), even if this solution exists. For example, it is instructive to consider a class of radiative cooling rates having the form
\beq
Q_{\rm rad}(n,T) \sim n^\alpha\, T^\beta \, .
\label{eq-Q_rad-powerlaw}
\eeq
This functional form actually reflects many real astrophysically-relevant cooling mechanisms such as  bremsstrahlung 
(corresponding to $\alpha = 2, \, \beta=1/2$) and cyclotron/synchrotron and inverse-Compton cooling ($\alpha=1, \, \beta=1$). 
With this functional form, the condition~(\ref{eq-evolutionary-condition}) becomes
\beq
\alpha < 1+\beta \, .
\label{eq-evolutionary-condition-powerlaw}
\eeq

Thus we see that in the case $\alpha=1,\, \beta=1$ corresponding to cyclo/synchrotron and external inverse Compton cooling, important in many high-energy astrophysical environments, the condition~(\ref{eq-evolutionary-condition}) is easily satisfied. This suggests that strong radiative cooling of reconnection layers by these mechanisms may be important, provided that other necessary conditions are also satisfied. 

On the other hand, thermal bremsstrahlung cooling ($\alpha = 2, \, \beta=1/2$) does not satisfy the above evolutionary condition. This means that either a large-$A$ solution exists but the system is not able to reach to it because 
the condition~(\ref{eq-SP-cooling-condition}) is not satisfied, or the condition~(\ref{eq-SP-cooling-condition}) is satisfied
and the reconnection layer starts to cool and compress rapidly but a stationary $A\gg 1$ solution does not exist and so the system experiences a catastrophic cooling collapse which can only be stopped by other processes, e.g., by the transition to the optically thick regime. (In that latter case, however, it is not clear how one would able to avoid a similar cooling catastrophe in the ambient plasma outside the reconnection layer.)

Moreover, the same conclusion can be generalized to any realistic emission process due to binary collisions between electrons and ions, including atomic line cooling, which is the predominant cooling process in the solar corona in the temperature range between about $10^5$ and $10^7$~K. Because of the collisional nature of these processes, their corresponding cooling rate is proportional to the square of the plasma density, $Q_{\rm rad}(n,T) = n^2 \, \Lambda(T)$ , 
as described by equation~(\ref{eq-Q_ff}).
Hence, $\alpha=2$ for any such process, and the condition~(\ref{eq-evolutionary-condition-powerlaw}) yields $\beta > 1$, i.e., the cooling function $\Lambda(T)$ would have to be rising faster than the temperature. This does not appear to be the case for a realistic atomic-line cooling function in the above temperature range.

This conclusion is important, because the fact that the solar corona does not satisfy this evolutionary condition 
is the main reason why prompt radiative cooling is not important in coronal reconnection events (e.g., solar flares). 
Indeed, one can easily see that a solution with large~$A$ does exist for solar coronal parameters; therefore, without the above evolutionary condition~(\ref{eq-evolutionary-condition}), one would expect to see instances of strongly radiatively cooled reconnection layers in solar flares, which would have dramatic observational consequences. The fact that we don't actually see strong radiative reconnection layers in the solar corona is unambiguously explained by the failure of the relevant cooling mechanisms to satisfy the condition~(\ref{eq-evolutionary-condition}).

A related very important issue is the {\it thermal stability} of the reconnection layer. 
Let us assume that a stationary solution $A=A_* \gg 1$ does indeed exist, $Q_{\rm rad}(A_*) = Q_{\rm ohm}(A_*) = 
A_* \, Q_0$. Then one can easily see that if the function $Q_{\rm rad}(A)/Q_{\rm ohm}(A) \sim Q_{\rm rad}(A)/A$ has a negative derivative at $A=A_*$, then this solution is stable. If, on the other hand, this function has a positive derivative at $A=A_*$, then the solution is unstable: a slight increase in~$A$ will make $Q_{\rm rad}$ larger than $Q_{\rm ohm}$, which will lead to a stronger compression so that $A$ will increase even further, etc. For the power-law form of the radiative cooling rate, equation~(\ref{eq-Q_rad-powerlaw}), we have $Q_{\rm rad}(A)/Q_{\rm ohm}(A) \sim A^{\alpha-\beta-1}$, and hence the condition that the layer is thermally stable is $\alpha<1+\beta$, which coincides with the above 
evolutionary condition~(\ref{eq-evolutionary-condition-powerlaw}). Thus, EIC leads to a stable reconnection layer,  whereas the bremssrtahlung solution is unstable. 


It is important to note that in many realistic situations several radiative mechanisms are acting simultaneously and, in particular, different cooling processes may dominate at different states of the reconnection layer, which may make the function $Q_{\rm rad}(A)/A$ non-monotonic. For example, it may be possible that one process, characterized by a decreasing function $Q_{\rm rad}(A)/A$ (such as EIC), is responsible for the initial radiative-cooling collapse of the hot, low-density, $A\sim 1$ reconnection layer to a cooler, denser, $A\gg 1$ state, at which point another process, e.g., bremsstrahlung, becomes more effective and takes over as the main cooling mechanism. If the cross-over value~$A_c$
at which $Q_{\rm rad}^{EIC}(A_c) = Q_{\rm rad}^{brems}(A_c)$ is higher than the value of~$A$ corresponding to the stable stationary EIC-cooling solution, see equation~(\ref{eq-EIC-A}), then that solution is reached before bremsstrahlung becomes effective and the above transition to the bremsstrahlung regime does not occur. 
In the opposite case, however, the total (EIC plus bremsstrahlung) cooling rate at the cross-over point~$A_c$ is higher than~$Q_{\rm ohm}(A_c)\simeq A_c\, Q_0$, and the system transitions to the bremsstrahlung cooling regime. Then, the catastrophic cooling collapse continues. Eventually, the plasma reaches such low temperatures and high densities that either the layer becomes optically thick or recombination becomes important and changes the overall physical picture. As follows from the above discussion, the condition for this two-stage, hybrid scenario to work can be formulated as 
\beq
Q_{\rm ohm}(A_c)  \sim A_c \, Q_0 < Q_{\rm rad}^{\rm tot}(A_c)  = 2\, Q_{\rm rad}^{EIC} \simeq  \tau_T\, {c\over{V_{A0}}}\, {U\over{n_0 m_e c^2}} \, Q_0\, , 
\eeq
where we used equation~(\ref{eq-Q_rad-EIC-cyclo-2}) to get the right-hand side, and where $A_c$ is the solution of the equation $Q_{\rm rad}^{EIC}(A) = Q_{\rm rad}^{brems}(A)$. Using expressions (\ref{eq-Q_ff-A}) and~(\ref{eq-Q_rad-EIC-cyclo-2}), we find that the cross-over takes place at
\beq
A=A_c = 
\biggl({\sqrt{3\pi}\over{2\sqrt{2}}}\, \bar{g}_{\rm ff}^{-1}\,\alpha_{\rm fs}^{-1}\, {V_{A0}\over{c}}\, \sqrt{m_p\over{m_e}}\, {U\over{n_0 m_e c^2}}\biggr)^{2/3} \, .
\eeq
Then, the condition for the transition to the bremsstrahlung-cooling collapse becomes 
\beq
{U\over{n_0 m_p c^2}} > {3\pi\over{8}}\, (\alpha_{\rm fs}\, \bar{g}_{\rm ff})^{-2}\,  \tau_T^{-3}\, \biggl({V_{A0}\over{c}}\biggr)^5 \, .
\label{eq-condition-hybrid}
\eeq

We believe this two-stage (hybrid) reconnection-layer cooling process may be important in some high-energy astrophysical environments, although probably not in the solar corona, where the EIC and cyclotron cooling mechanisms are not effective.


\subsection{Collisional Reconnection Condition}
\label{subsec-collisional}

Throughout this paper we have assumed that we are dealing with a resistive-MHD, collisional reconnection process. 
In the absence of a strong guide field, a commonly used condition for validity (for non-relativistic electron-ion plasma) of this assumption can be cast as~\citep[e.g.,][]{Ma_Bhattacharjee-1996, Cassak_etal-2005, Yamada_etal-2006, Uzdensky-2007a,  Uzdensky-2007b, Uzdensky-2007c}
\beq
\delta_{\rm SP} > d_i \, , 
\label{eq-condition-collisional}
\eeq
where 
\beq
d_i \equiv {c\over{\omega_{pi}}} = \sqrt{m_i\over{m_e}}\, {1\over{\sqrt{4\pi n r_e}}}
= \sqrt{m_p\over{m_e}}\, {1\over{\sqrt{4\pi n_0 r_e}}} \, A^{-1/2} 
\label{eq-def-d_i}
\eeq
is the so-called ion collisionless skin depth ($r_e\equiv e^2/m_e c^2$ is the classical electron radius). If the condition~(\ref{eq-condition-collisional}) is not satisfied, then the Sweet-Parker slow resistive reconnection is not applicable and one may have a much faster collisionless reconnection regime (attributed to either Hall effect or anomolous resistivity). 

Using equation (\ref{eq-delta_of_A}), we have 
\beq
{{\delta^2}\over{d_i^2}} \sim 
{3\over 2}\, C_\eta\, \ln\Lambda\, \tau_T\, {m_e\over{m_p}}\, {c\over{V_{A0}}}\, \theta_e^{-3/2} \simeq
12\, C_\eta\,  \ln\Lambda\, \tau_T\, \biggl({m_e\over{m_p}}\biggr)^{5/2} \, \biggl({V_{A0}\over{c}}\biggr)^{-4} \,  A^{3/2} \, .
\eeq

Thus, the condition that the layer is in the collisional regime, $\delta> d_i$, can be cast as
\beq
\tau_T\, A^{3/2} > {1\over{12 C_\eta\, \ln\Lambda}}\, \biggl({m_p\over{m_e}}\biggr)^{5/2}\, 
\biggl({V_{A0}\over{c}}\biggr)^4 \, .
\eeq

More specifically, in the bremsstrahlung case, using equation~(\ref{eq-brems-A}) 
we see that the requirement that the layer is collisional reads 
\beq
{V_{A0}^2\over{c^2}}  > {\sqrt{2}\over{18\pi\sqrt{3\pi}\,C_\eta}}\, 
{{\alpha_{\rm fs}^3\, \bar{g}_{\rm ff}^3}\over{\ln\Lambda}}\, \tau_T^2 \, {m_p\over{m_e}}  = 
 {1\over{6\sqrt{3}\pi}}\, {{\alpha_{\rm fs}^3\, \bar{g}_{\rm ff}^3}\over{\ln\Lambda}}\, \tau_T^2 \, {m_p\over{m_e}} 
 \simeq 2\times 10^{-5}\, {{\bar{g}_{\rm ff}^3}\over{\ln\Lambda}}\, \tau_T^2\, , 
\eeq
which is easily satisfied for many astrophysical plasmas.

Similarly, for external inverse Compton cooling, using equation~(\ref{eq-EIC-A}), the collisional reconnection condition becomes
\beq
\biggl({{8\pi\, U}\over{B_0^2}}\biggr)^{3/2}  > 
{2\over{3C_\eta\, \ln\Lambda}}\, {{m_p}\over{m_e}}\, \tau_T^{-5/2}\, \biggl({{V_{A0}}\over{c}}\biggr)^{5/2} \, .
\eeq


\subsection{Optically-Thin Condition}
\label{subsec-opt-thin}

In this paper we have restricted our analysis to the simplest case of an optically thin reconnection layer. 
The reason for this simplification was that we wanted to avoid the complexities associated with the full 
radiative-transfer problem that one has to consider in the more general optically thick or intermediate case.

The basic necessary condition for the validity of the optically-thin approximation is the requirement that the Thomson optical depth across the layer, $\tau_T(\delta)$, be small. Using equation~(\ref{eq-delta_of_A}), we can express this optical depth in terms of~$A$ for the case of Spitzer resistivity as
\beq
\tau_T(\delta) \sim \delta \sigma_T n \sim \tau_T\, A\, {\delta\over{L}} \sim \tau_T \, A^{1/2}\, S_0^{-1/2}  \sim 
\tau_T\, A^{5/4}\, \biggl({{V_{A,0}}\over{c}}\biggr)^{-2} \, \biggl({m_e\over{m_p}}\biggr)^{3/4}\, 
\sqrt{{8 r_e\, \ln\Lambda}\over{L}} \, .
\eeq

Because the layer length~$L$ is so much larger than~$r_e\sim 3\times 10^{-13}\, {\rm cm}$ 
(by many many orders of magnitude) for the absolute majority of potential applications of our theory, 
we see that this optical depth is essentially always much smaller than~1, as we have assumed.

However, we have to be mindful that Thomson scattering is not the only source of opacity. 
In particular, at high densities and low temperatures, free-free absorption, proportional to~$n^2 T^{-7/2}$, 
may dominate over the electron scattering. In addition, in the relativistic case with cyclo/synchrotron cooling, 
self-absorption may sometimes become important.


\subsection{Electron/Ion Temperature Equilibration Condition}
\label{subsec-T_e=T_i}

In this paper we have also implicitly assumed that the electron and ion temperatures are equal, or, at least,
that the ion temperature is not much higher than the electron temperature. If this were not the case, i.e., if we had
$T_i\gg T_e$, then our pressure balance condition would have been significantly modified; namely, a pressure balance would be established between the magnetic pressure outside the layer and the {\it ion} pressure inside the layer. As a result, 
one would not have been able to use the pressure balance to determine the {\it electron} temperature in terms of~$A$, 
and it is the electron temperature that is needed in order to estimate the plasma resistivity and the cooling rate. 

Thus, in order to delineate the realm of validity of our model more accurately, we need to figure out when the approximation $T_e\simeq T_i$ holds. To do this, we follow the approach of~\cite{Uzdensky-2007c}. The electron-proton temperature equilibration time $\tau_{\rm EQ}$ is by a factor $m_p/m_e$ longer than the electron-ion momentum-transfer collision time~$\tau_{ei}$; expressing the latter in terms of the electron collisionless skin depth and the collisional magnetic diffusivity, $\tau_{ei} \sim d_e^2/\eta$, we obtain $\tau_{\rm EQ} \sim d_i^2/\eta$. It it convenient to recast this estimate using expression~(\ref{eq-delta}) as follows:
\beq
\tau_{\rm EQ} \sim {{d_i^2}\over{\eta}} \sim \tau_{A0}\, A^{-1}\, {{d_i^2}\over{\delta^2}} \, 
\eeq

Because we are interested in the strong-compression regime, $A\gg 1$, we see that the condition that electrons and ions 
are thermally well coupled, $\tau_{\rm EQ} \ll \tau_{A0} $ is in fact weaker than the condition that the layer is in the collisional regime, $\delta> d_i$.


\section{Strong Guide-Field Case}
\label{sec-guide}

In the presence of a non-zero guide (i.e., out of plane) magnetic field component~$B_z$, the situation is somewhat more complicated, in part because one can no longer use the cross-layer pressure balance to determine straight-forwardly 
the central temperature in terms of the compression ratio. The guide field provides a certain pressure that resists compression. The pressure balance across the layer with an upstream reconnecting magnetic field~$B_0$, and an upstream guide field~$B_{z0}$ can be written as (again, neglecting the upstream thermal pressure):
\beq
{{B_0^2}\over{8\pi}} + {{B_{z0}^2}\over{8\pi}} = {{B_{zc}^2}\over{8\pi}} + P_c \, . 
\eeq
where $B_{zc}$ and $P_c = 2 n k_B T_c$ are the values of the guide field and the plasma pressure, respectively, at the center of the layer.

Assuming that the guide field compresses in proportion to the gas density, i.e., 
\beq
{B_{zc}\over{B_{z0}}} = {n\over{n_0}} = A \, , 
\eeq
we can then write the above pressure balance as
\beq
P_c = 2 A\, n_0\, k_B T_c = {{B_0^2}\over{8\pi}}  + {{B_{z0}^2}\over{8\pi}}\, (1-A^2) \, ,
\label{eq-pressure-balance-guide}
\eeq
which represents a relationship between $T_c$ and~$A$ that is somewhat more complicated than the simple relationship~(\ref{eq-T_of_A}) that we obtained in zero-guide field case. In order to close the problem, we need to use the heating/cooling balance to get a second relationship between $T_c$ and~$A$. In the most interesting for us strong-cooling regime, this balance is basically the same as in the zero-guide field case and is given by expression~(\ref{eq-heat-cool-1}):
\beq
Q_{\rm rad}(An_0, T_c) = Q_{\rm ohm}  \sim A \, Q_0  \, .
\label{eq-heat-cool-guide}
\eeq

Equations (\ref{eq-pressure-balance-guide}) and~(\ref{eq-heat-cool-guide}) represent a closed system of two algebraic equations for the unknowns, $A$ and~$T_c$, and thus, in principle, completely determine the thermodynamic structure of the radiatively-cooled reconnection layer with a finite guide field. 

One particularly simple limit is the case of a strong guide field, $B_{z0} \gg B_0$. In this case, the guide field essentially provides a high uniform background pressure and hence effectively enforces incompressibility. Then the pressure balance equation~(\ref{eq-pressure-balance-guide}) and the heating/cooling balance equation~(\ref{eq-heat-cool-guide}) decouple from each other. Namely, in the pressure balance equation the gas pressure can be neglected and so the temperature drops out; this equation then only involves the compression ratio~$A$ and yields a trivial result $A=1$ (to lowest order in $B_0/B_{z0}$). Then, the remaining heating/cooling equation~(\ref{eq-heat-cool-guide}) becomes simply
\beq
 Q_{\rm rad}(n_0, T_c) =  Q_{\rm ohm} = Q_0 = {{B_0^2}\over{4\pi\, \tau_{A0}}} \, ,
\label{eq-heat-cool-guide-strong}
\eeq
and one can use it to determine the layer's temperature and therefore the parallel Spitzer resistivity 
$\eta_\parallel \simeq \eta_\perp/2$ and the corresponding Lundquist number~(\ref{eq-Spitzer-S}):
\beq
S_\parallel \simeq 2\, S_\perp \simeq
{2\over{C_\eta}}\, {L\over{r_e\, \ln\Lambda}}\, {{V_{A0}}\over c}\, \theta_e^{3/2}\, .
\label{eq-S-guide}
\eeq

Then, the reconnection rate and the thickness of the layer are then given by the usual Sweet--Parker scalings: 
\beq
{v_{\rm rec}\over{V_{A0}}} \sim {\delta\over L} \sim S_\parallel^{-1/2} \simeq 
\sqrt{{{C_\eta}\over 2}\, {{c\,r_e\, \ln\Lambda}\over{V_{A0}\, L}}}\, \theta_e^{-3/4} \, .
\label{eq-rec_rate-guide}
\eeq

For example, in the case of bremsstrahlung, $Q_{\rm rad} = n^2\, \kappa T^{1/2}$ 
[see eqs.~(\ref{eq-Q_ff}) and~(\ref{eq-Lambda-brems})-(\ref{eq-kappa-brems})], we readily obtain
\beq
\theta_e^{1/2} \simeq \sqrt{3\pi\over 8}\, {1\over{\alpha_{\rm fs}\, \bar{g}_{\rm ff}\, \tau_T}}\, 
\biggl({V_{A0}\over{c}}\biggr)^3\, {m_p\over{m_e}} \, ,
\eeq
and hence
\beq
{v_{\rm rec}\over{V_{A0}}} \sim {\delta\over L} \sim {4\over{3\pi\, 3^{1/4}}} \, \sqrt{{r_e\, \ln\Lambda}\over{L}}\, (\alpha_{\rm fs}\, \bar{g}_{\rm ff}\, \tau_T)^{3/2}\, \biggl({{V_{A0}}\over c}\biggr)^{-5}\, \biggl({m_e\over{m_p}}\biggr)^{3/2} \, .
\eeq

In the case of EIC cooling, $Q_{\rm rad} = 2\, \sigma_T c \, U \, n \, \theta_e$  
[see eq.~(\ref{eq-Q_rad-EIC-cyclo})], we get
\beq
\theta_e \simeq {{B_0^2}\over{8\pi\, U}}\, {V_{A0}\over{c}}\, \tau_T^{-1} \, , 
\label{eq-EIC-theta-guide}
\eeq
which, interestingly, coincides with the result~(\ref{eq-EIC-theta}) that we have obtained for the zero-guide field case. 
The main reason for this clearly lies in the fact that, apart from its temperature dependence, the EIC radiative cooling rate~(\ref{eq-Q_rad-EIC-cyclo}) scales with the compression ratio~$A$ in the same way (linearly) as the ohmic heating rate $Q_{\rm ohm} \sim A\, Q_0$. Therefore, the $A$ factors on both sides of the heating/cooling balance cancel out and the resulting temperature turns out to be independent of the plasma compression ratio (which is controlled by~$B_z$).

In the case of cyclotron cooling, the situation is very similar, we just need to replace $U$ in the expression for the radiative cooling rate by the magnetic energy density, $U\simeq U_{\rm mag} \simeq B_{z0}^2/8\pi$. From equation~(\ref{eq-heat-cool-guide-strong}) we then immediately get: 
\beq
\theta_e \simeq {{B_0^2}\over{B_{z0}^2}}\, {V_{A0}\over{c}}\, \tau_T^{-1} \, . 
\label{eq-cyclo-theta-guide}
\eeq
which differs from the corresponding zero-guide-field expression only by a factor of~$B_0^2/B_{z,0}^2$.

Recall that, as we discussed in~\S~\ref{subsec-EIC-Cyclotron}, in the $B_z=0$ case the condition that the electrons inside the reconnection layer are non-relativistic ($\theta_e \ll 1$) coincides with the condition $\Omega_{ce}<\omega_{pe}$ (because of the pressure balance and the $T_e=T_i$ assumption). Then, the cyclotron photons cannot propagate effectively through the layer and this is why we chose to neglect the cyclotron cooling process for that (relativistic zero-guide field) case. We would like to note, however, that in the case of a strong guide field, the restrictions due to the model assumptions appear to be not as severe; in particular, there exist a parameter regime where cyclotron cooling can be quite effective. This is possible because now the electron temperature is no longer determined from the pressure balance and can in fact be quite low. Then, the electron cyclotron frequency $\Omega_{ce}$ can be higher than~$\omega_{pe}$ provided that $B_{z0}^2 > 4\pi \, n_0 m_e c^2$. 

We thus see that a more powerful soft radiation field or a stronger magnetic guide field lead to a more efficient inverse-Compton or cyclotron cooling, respectively, which results in a lower plasma temperature. This, in turn, makes the reconnection process go faster because of the~$T^{-3/2}$ dependence of the Spitzer resistivity on the temperature. 
In particular, from equation~(\ref{eq-rec_rate-guide}), we get
\beq
{v_{\rm rec}\over{V_{A0}}} \sim {\delta\over L} \sim 
\sqrt{{{C_\eta}\over 2}\, {{r_e\, \ln\Lambda}\over{L}}}\, \biggl({{V_{A0}}\over c}\biggr)^{-5/4}\, 
\biggl(\tau_T \, {{8\pi\, U}\over{B_0^2}}\biggr)^{3/4} \, ,
\eeq
where, once again, $U = U_{\rm rad} + B_{z0}^2/8\pi$.


\section{Conclusions}
\label{sec-conclusions}


In this paper we presented a theoretical analysis of magnetic reconnection in the presence of a strong radiative cooling. Our study was motivated by the increasing interest in reconnection in the high-energy astrophysics community and in the laboratory high-energy-density community (e.g., laser-plasma and z-pinch communities). Both these areas of research deal with environments in which the energy density is so high that various radiative effects (radiative cooling being just one of them) become important.

We would like to stress that this situation can be contrasted with traditional studies of magnetic reconnection. 
These studies are usually motivated by applications to relatively tenuous, low-energy-density solar-system environments,
such as the solar corona, the Earth magnetosphere, laboratory magnetic fusion devices such as tokamaks,  and also dedicated laboratory reconnection experiments such as the~Magnetic Reconnection Experiment~\citep{Yamada_etal-1997}. One can easily show that in all these systems the effects of radiation on the reconnection process are negligible, and this fact justifies the complete omission of radiation from the traditional reconnection models. This means, however, that neither any concrete results, nor general physical insights from these studies can be applied  with any degree of confidence to reconnection in high-energy-density environments where radiative effects are important. 
We thus believe that investigating the effects of radiation on magnetic reconnection is an important new area of research with potentially significant applications to astrophysics. Since this area has been essentially unexplored so far, it should be viewed as a new frontier in reconnection research. 

The physics involved in reconnection with radiation is rather complex, as there are several different physical effects that radiation can exert, e.g., radiative cooling, radiation pressure, radiative resistivity (photon drag), and, in the most extreme astrophysical cases, pair creation. We of course do not hope to cover all this complexity in one paper and instead take a step-by-step approach. We therefore view the present paper as being a first in a series, and so we restricted this study to what we think is the simplest and most fundamental first step --- non-relativistic,  steady-state, collisional, resistive-MHD reconnection in 2D with classical Spitzer resistivity and with radiative cooling. We thus left the above-mentioned other potentially important radiative effects for a future study. 

Our analysis in this paper basically followed the classical Sweet--Parker approach but added thermodynamic considerations, in particular, the balance between ohmic heating and radiative cooling. 
We devoted most of the paper to the case of strictly anti-parallel reconnection, with zero guide magnetic field. 
We showed that, because of the need to maintain the pressure balance with the outside magnetic field pressure, 
strong radiative cooling inevitably leads to a strong compression of the plasma inside the reconnection layer, $A\equiv n/n_0\gg 1$, where $n$ is the density inside the layer and~$n_0$ is the upstream density just outside the layer.
We then demonstrated that, in contrast to the incompressible Sweet--Parker case, magnetic tension dominates over the pressure gradient as the main force accelerating the plasma outflow from the layer and that the resulting outflow velocity is of order the Alfv\'en speed in the {\it uncompressed}, upstream plasma. Then, following a Sweet--Parker-like analysis, we derived scalings for the the reconnection velocity~$v_{\rm rec}$ and the thickness~$\delta$ of the layer and showed that 
the former is faster by a factor of~$A^{1/2}$, and the latter is thinner by the same factor, relative to the corresponding classical Sweet--Parker expressions without radiative cooling. 

We then considered the particular case of Spitzer resistivity and found, not surprisingly, that there was another important effect that radiative cooling exerts on the reconnection process. Namely, strong cooling makes the layer temperature lower relative to the case without cooling, and this, in turn, leads to higher Spitzer resistivity and, correspondingly, a lower Lundquist number, which results in faster reconnection. 

We then argued that in order to find the equilibrium layer temperature and the corresponding compression ratio, one needs to use the condition of balance between the ohmic heating and radiative cooling (since in the strong cooling regime the advective heat losses can be neglected), and we derived a general equation expressing this balance. 
The key physical idea here is that, whereas in the Sweet--Parker model the matter and the energy go together (both advected out along the layer), in the strong radiative cooling case they partly separate: while the matter is still advected out of the layer, a significant part of the released energy is promptly radiated away across the layer.

We illustrated these ideas by considering several astrophysically important cooling mechanisms:
bremsstrahlung and cyclotron and external inverse Compton emission. For simplicity, we restricted the analysis in this paper to the optically-thin case, leaving the optically-thick and a general intermediate cases to a future study. We derived specific expressions for the key reconnection parameters for each of these processes. 

We then considered the restrictions that our various assumptions imposed on the model, such as the condition that the layer is in the collisional resistive-MHD regime ( $\delta > d_i$) and the condition that it is optically thin. Furthermore, we showed that in addition to the $A\gg 1$ strong compression condition, one must impose an evolutionary condition for the system to be able to reach the stationary strong-cooling solution starting from its initial hot state. We also showed that, related to that condition, there is also a condition of thermal stability of the layer. For a cooling rate with a general power-law form, $Q\sim n^\alpha\, T^\beta$, these conditions both become simply $\alpha< 1+ \beta$. Interestingly, bremsstrahlung (as well as the solar coronal cooling function dominated by various atomic lines) does not satisfy this condition. We believe this means that, depending on the particular parameters of a given system, bremsstrahlung is either ineffective or leads to a thermal cooling catastrophe. On the other hand, the EIC  mechanism does satisfy this condition and hence may lead to a stable stationary strong-cooling solution. There may also be a hybrid scenario where the initial EIC cooling stage is followed by a bremsstrahlung-dominated cooling collapse. 

In the last section of our analysis we discussed the effect of radiative cooling on reconnection with a non-zero guide field, and, in particular, concentrated on the case with a strong guide field. We again assumed the Spitzer resistivity and derived specific analytical expressions for the reconnection rate and the layer thickness. We found that, because a strong guide field effectively enforces incompressibility, there is no acceleration of reconnection due to the plasma compression that one has in the zero guide-field case. Thus, the only way in which radiative cooling affects (namely, accelerates) the reconnection process in the strong guide field case is through its effect on the plasma temperature and then through it --- on the Spitzer resistivity and the Lundquist number. 

To sum up, in this paper we made the first strides towards understanding the role of radiation (namely, optically thin radiative cooling) in high-energy density magnetic reconnection. We hope that, fueled by its potential astrophysical and laboratory high-energy-density applications, this important area of research will grow in the coming years.

\acknowledgments

\begin{acknowledgments}


This work is supported in part by National Science Foundation Grant  PHY-0903851 (DAU)
and by NASA Chandra Fellowship PF7-80048 (JCM).

\end{acknowledgments}




\bibliographystyle{apj}

\bibliography{xxx}

\clearpage


\end{document}